\documentstyle[11pt]{article}

\newcommand{\be}{\begin{equation}}
\newcommand{\ee}{\end{equation}}
\newcommand{\bea}{\begin{array}}
\newcommand{\ea}{\end{array}}
\newcommand{\beqa}{\begin{eqnarray}}
\newcommand{\eeqa}{\end{eqnarray}}
\newcommand{\bean}{\begin{eqnarray*}}
\newcommand{\eean}{\end{eqnarray*}}

\def\Rtwo{\matrix{{}\cr  R \cr {}^2}}
\def\Rone{\matrix{{}\cr  R \cr {}^1}}

\def\BI{{\rm 1\!l}}
\def\vonex{\matrix{{}\cr v(x) \cr {}^1}}
\def\vtwox{\matrix{{}\cr v(x) \cr {}^2}}
\def\vtwoy{\matrix{{}\cr v(y) \cr {}^2}}
\def\gonex{\matrix{{}\cr g(x) \cr {}^1}}
\def\gtwoy{\matrix{{}\cr g(y) \cr {}^2}}

\def\eonei{\matrix{{}\cr e^i \cr {}^1}}

\def\up#1{\leavevmode \raise.16ex\hbox{#1}}

\newcommand{\journal}[4]{{\sl #1 }{\bf #2} \up(19#3\up) #4}

\newcommand{\gapproxeq}{\lower .7ex\hbox{$\;\stackrel{\textstyle >}{\sim}\;$}}
\newcommand{\lapproxeq}{\lower .7ex\hbox{$\;\stackrel{\textstyle <}{\sim}\;$}}

\newcounter{appendice}

\def\thebibliography#1{{\bf REFERENCES\markboth
 {REFERENCES}{REFERENCES}}\list
 {[\arabic{enumi}]}{\settowidth\labelwidth{[#1]}\leftmargin\labelwidth
 \advance\leftmargin\labelsep
 \usecounter{enumi}}
 \def\newblock{\hskip .11em plus .33em minus -.07em}
 \sloppy
 \sfcode`\.=1000\relax}

\begin{document}

\begin{flushright}
UAHEP 9811\\
Nov 1998\\
\end{flushright}

\centerline{ \LARGE  Hamiltonian Approach to Poisson Lie T-Duality}
\vskip 2cm

\centerline{ {\sc  A. Stern }  }

\vskip 1cm

\centerline{  Dept. of Physics and Astronomy, Univ. of Alabama,
Tuscaloosa, Al 35487, U.S.A.}

\vskip 2cm

\vspace*{5mm}

\normalsize
\centerline{\bf ABSTRACT}
The Hamiltonian formalism offers a natural framework
for discussing the notion of Poisson Lie T-duality.
This is because the duality is inherent in the
Poisson structures alone and exists regardless of the
 choice of  Hamiltonian.
Thus one can pose alternative dynamical systems possessing
nonabelian    T-duality.     As an example,
we find a dual Hamiltonian formulation of the
$O(3)$ nonlinear $\sigma$-model.   In addition,
 we easily recover the known dynamical systems having
 Poisson Lie T-duality
 starting from a general quadratic Hamiltonian.

\vskip 2cm
\vspace*{5mm}

\newpage
\scrollmode

\section{Introduction}

Poisson Lie T-duality \cite{ks95},\cite{ksb376},\cite{tyun},\cite{akt}
  is a nonabelian generalization of Abelian
T-duality which shares many structural features with the Abelian
system.
  Unlike in other  approaches\cite{prv}, it does not require the
existence of an isometry group.  The formulation of Poisson Lie T-duality
 given so far is tied to an action principle.  It therefore
  implies a particular
dynamics, which is namely that of  a $\sigma$-model on a group manifold
  in a background
obeying certain geometrical criteria.  Thus
duality  links  not only
different backgrounds,  but different target space manifolds,
corresponding to different Lie groups.

The associated Hamiltonian descriptions of the dual
systems were derived recently by Sfetsos\cite{Sfet}.
There it was shown that the transformations
between  systems
are canonical transformations.  Therefore in the Hamiltonian framework,
one sees that  Poisson Lie T-duality
 can  be introduced independently of the dynamics, as it is
a feature of the Poisson algebra alone.  One should then be able to
 pose alternative dual dynamical systems.
We wish to explore that possibility here.

  For   Poisson Lie T-duality one introduces the Drinfeld double group
   $D$, which by definition has a pair $(G,H)$ of maximally isotropic
     subgroups\cite{drin}.
  The relevant Poisson algebra for the theory  is $\widehat{LD}$,
   the central extension of the loop  group of $D$.\cite{Sfet}
    This algebra can be
  realized on either   $LT^*G$ or $LT^*H$,
    the loop algebras of $T^*G$ or  $T^*H$, respectively.
     For that reason one can consider
   $LT^*G$ and $LT^*H$ as being `dual'.
The  duality will persist for any Hamiltonian dynamics we
 introduce on $\widehat{LD}$.     Arbitrary choices for the
  Hamiltonian may, in general,
 lead to exotic dynamics, which need not  be Lorentz invariant or local.
On the other hand, as we shall show,
by restricting to quadratic Hamiltonians ${\tt H}$ on  $\widehat{LD}$
 and demanding Lorentz
invariant dynamics, more specifically, the dynamics of nonchiral scalar
 fields,  one is able to recover the known systems.
The dual  Hamiltonians derived in \cite{Sfet} are then recovered by
projecting ${\tt H}$ onto    $LT^*G$ and $LT^*H$.

Concerning alternative dynamical systems, three
 approaches can be explored:
1) Instead of utilizing the background fields on $G$ given
 in \cite{ks95},
   one is free to choose the  backgrounds arbitrarily.
In general, however, it may not be possible to  simultaneously have
 local  expressions for the  backgrounds on both $G$ and $H$.
2)  Instead of restricting to quadratic Hamiltonians, one can examine
the possibility of  higher order terms.
3)  Gauge symmetries can  be  introduced by imposing
first class constraints.  The latter must generate a
subgroup of $\widehat{LD}$.  The target manifolds are now  coset spaces,
 and thus one ends up with dual Hamiltonian
 descriptions of coset models.    We will briefly examine this last
  approach for the case  of the $O(3)$ nonlinear $\sigma$-model.

The outline for this article is as follows:
In Sec. 2, we give a simple construction of the duality between
Poisson algebras  $LT^*G$ and $LT^*H$.  In Sec. 3, we examine a general
quadratic Hamiltonian  on   $\widehat{LD}$, and give
its projection onto $LT^*G$ and $LT^*H$.
  We then demonstrate how to recover
the known dynamics for the Abelian and nonabelian cases in
Sec's 4 and 5.  We examine the case of the
 $O(3)$ nonlinear $\sigma$-model in Sec. 6.

\section{Duality of Poisson algebras}
\setcounter{equation}{0}

 In defining   the Drinfeld double group
    $D$, and its maximally isotropic subgroups $G$ and $H$,
say  $G$ and $H$ are   $n$ dimensional
  generated by $e_i$  and $ e^i$, $ i=1,2,....n$,
 respectively, their union generating all of $D$.
   By definition the  commutation relations
     and invariant scalar product can be written in the form\cite{drin}
 \beqa [e_i,e_j]&=&c_{ij}^k e_k \cr
 [ e^i, e^j]&=& c^{ij}_k  e^k \cr
 [ e^i,e_j]&=&c_{jk}^i  e^k  -  c^{ik}_j e_k \;, \label{dbalg}
 \eeqa
  $$< e^i|e_j>=\delta_j^i\;,\qquad < e^i| e^j>=   <e_i|e_j>=0 \;.$$

The  phase space of interest is spanned by
  fields $v(x)=e^iv_i(x)+e_iv^i(x)$     taking values
in the Lie-algebra associated with $D$, $x$ being coordinates of
 a one dimensional spatial domain.
As stated in the introduction, the relevant Poisson algebra is
the central extension of the loop  group of $D$\cite{Sfet}:
\be   \{\vonex,\vtwoy\} = [C,\vonex]\delta(x-y) - C\partial_x \delta(x-y)
\;,\label{lgoD}\ee
where we  use tensor product notation,
the $1$ and $2$ labels referring to two separate vector spaces, with
$\vonex=v(x)\otimes \BI$,  $\vtwoy=\BI\otimes v(y)$, and
$\BI$ being the unit operator acting on the vector spaces.
$C$ in (\ref{lgoD}) is a constant adjoint invariant tensor and hence
$[C,\vonex+\vtwox]=0$.  We normalize it according to:
 $C=e^i \otimes e_i + e_i \otimes e^i \;.$

The algebra (\ref{lgoD}) can
be realized on either $LT^*G$ or $LT^*H$,
giving two dual descriptions of the phase space.
  The former is  spanned by the fields $g(x)$ and $J_i(x)$,
$g(x)$ taking values in $G$, while  $J_i(x)$
 generate left translations on $G$, i.e.
\beqa   \{ J_{i} (x),  g (y) \}& =&e_i g(x) \delta(x-y) \cr
\{ J_{i} (x), J_{j} (y) \}&=&-c_{ij}^k  J_{k} (x)  \delta(x-y)
\;.\label{ltoG}\eeqa
In terms of these variables, $v$ is given by
 \be v=v(g,J_i)= g^{-1}e^i g J_i  +     g^{-1}\partial_x   g  \;.
\label{voG} \ee
 A straightforward calculation using  (\ref{ltoG})and (\ref{voG}) yields
 \be   \{\vonex,\vtwoy\} = \gonex^{-1}\gtwoy^{-1}\biggl(
  [C,\eonei] J_i(x)\delta(x-y) - C\partial_x \delta(x-y) \biggr)
  \gonex \gtwoy   \;.\ee
 To arrive at (\ref{lgoD}) one can simply apply the identity
\be  \gonex^{-1}\gtwoy^{-1}C\gonex \gtwoy \partial_x \delta(x-y)
=C \partial_x \delta(x-y)  +[\gonex^{-1}\partial_x
\gonex,C]\delta(x-y)
 \;.\ee

 This procedure can be repeated to  obtain the dual
 description associated with  $LT^*H$.
   The latter is   spanned by the fields $h(x)$ and $\tilde J ^i(x)$,
 $h(x)$ taking values in $H$, while  $\tilde J^i(x)$
  generate left translations on $H$.  The analogue of (\ref{ltoG}) is
 \beqa   \{\tilde J^{i} (x), h (y) \}& =&e^i h(x) \delta(x-y) \cr
 \{\tilde J^{i} (x),\tilde J^{j} (y) \}&=&-c^{ij}_k  \tilde J^{k} (x)
 \delta(x-y) \;.\label{dltoG}\eeqa
 In terms of these variables, $v$ is given by
 \be v=v(h,\tilde J^i)= h^{-1}e_i h \tilde J^i +h^{-1}\partial_x h \;.
 \label{voH}\ee

From the above it is evident that $LT^*G$ and $LT^*H$ have a common
 sector, i.e.  $\widehat{LD}$.
 Topology clearly prevents us, in general,
 from making a complete identification
of the dual descriptions.

For the case where both $G$ and $H$, the duality transformation has a
simple form.  If  we parametrize $g\in G$
 and $h\in H$ according to $g=\exp(e_i \phi^i)$ and $h=\exp(e^i \chi_i)$,
we get a nonlocal map between the canonically conjugate field variables
$(\phi^i, J_i )$ and their duals  $(\chi_i,\tilde J^i )$:
\beqa         J_i(x)& \rightarrow &\partial_x \chi_i(x)       \cr
\partial_x  \phi^i(x) &\rightarrow & \tilde J^i(x)   \;,\eeqa
which has the effect of interchanging winding modes with momentum
modes.\cite{kysy}
For the case of nonabelian $G$ and $H$, if we write
\beqa ge_i g^{-1}&=& a(g)_i^{\;\;j}e_j \cr
g e^i g^{-1}&=&b(g)^{ij} e_j +a(g^{-1})_j^{\;\;i} e^j\;,\label{defab}
 \eeqa   and
\beqa he^i h^{-1}&=& \tilde a(h)^i_{\;\;j}e^j \cr
h e_i h^{-1}&=&\tilde b(h)_{ij} e^j + \tilde a(h ^{-1})^j_{\;\;i} e_j\;,
\label{deftab}  \eeqa
which follow from the Lie algebra (\ref{dbalg}), a
duality transformation amounts to making the replacement:
\beqa   a(g)_i^{\;\;j}J_j &\rightarrow &  \tilde b(h^{-1})_{ij} \tilde J^j
+(h^{-1}\partial_xh)_i \cr  b(g^{-1})^{ij} J_j + (g^{-1}\partial_xg)^i
 &\rightarrow &   \tilde a(h)^i_{\;\;j}\tilde J^j\;. \eeqa

  In general,  local descriptions in one
 theory can  become nonlocal in the other.
To avoid nonlocal terms in the Hamiltonian ${\tt H}$ after projecting
onto  $LT^*G$ or $LT^*H$, below we will restrict to
 the case where ${\tt H}$ is a (local) function on  $\widehat{LD}$.
Starting from  general quadratic systems,  we shall show how
 the dynamics for free massless scalar fields  can be recovered
for the case of  Abelian T-duality,
 and also  how  the dynamical system
of Klimcik and Severa\cite{ks95} can be recovered in the nonabelian case.

\section{Dual Dynamics}
\setcounter{equation}{0}

To recover known systems, we examine
 Hamiltonians which are quadratic in $v$.
   The most general such Hamiltonian is
   \be {\tt H}=\frac12\int dx \; < v | Rv>\;,\label{Hvpvm}\ee
where $R$, for the moment, is an arbitrary linear operator.
It is  easy to compute the resulting Hamilton equations.
  Using  (\ref{lgoD}), we get
\beqa
\{\vonex,{\tt H}\} &=&\int dy  \; < \{\vonex,\vtwoy\}| \Rtwo \vtwoy >_2 \cr
&=&-<[C,\vtwox ]|\Rtwo \vtwox >_2 \;-\; <C|\Rtwo \partial_x\vtwox >_2 \cr
 &=&<C|\;[ \Rtwo \vtwox, \vtwox]  \;-\;  \Rtwo \partial_x\vtwox >_2 \cr
 &=& [\Rone \vonex, \vonex] \;-\;  \Rone \partial_x\vonex     \;,\eeqa
where $<\;,\;>_2$  denotes  a scalar product with  respect
 to the second vector     space in the tensor product.  Thus
\be\partial_t v = -R\partial_x v + [Rv,v]  \label{geom}  \;,\ee
which is the Muarer-Cartan equation on $D$ once we
  identify $v(x)$ and $-Rv(x)$ with the space  and time
components, respectively, of a one form $V$ on a two dimensional
space-time domain.    Then, at least on an open coordinate patch of the
domain, we can
write   $V=\ell^{-1}d\ell$, $\ell\in D$.  This yields
     \be \ell^{-1}\partial_t\ell + R(\ell^{-1}\partial_x\ell)
  =0 \;. \label{defpld}   \ee

These equations of motion have the same form as the defining equations
of  Klimcik and Severa \cite{ks95}.
  The  only difference is that we have not yet specified $R$.
Moreover, it was found \cite{ksb376},\cite{tyun},\cite{akt}  that
the  dynamical system  of  \cite{ks95}  could
be obtained starting from an action principle on $D$.
  The action corresponds
to the Wess-Zumino-Witten model (written on the  light cone), minus the
 term   (\ref{Hvpvm}), with a particular definition of the linear
operator $R$ .
The Wess-Zumino-Witten model (written on the  light cone) defines
 the symplectic two-form for the theory.  It corresponds precisely
to the Poisson algebra (\ref{lgoD}).  The remaining term then defines
the Hamiltonian  and thus also agrees with what we wrote (except for
us, $R$ is arbitrary).

 Without any loss of generality, we can write $R$ such that
\beqa  Re_i&=&- {\cal F}^0_{ij} e^j - {\cal H}^{0\;\;j}_{\;\;i} e_j \cr
  Re^i&=&-  {\cal H}^{0\;\;i}_{\;\;j} e^j -{\cal G}^{0ij} e_j \;,\eeqa
   where
${\cal F}^0$, ${\cal G}^{0}$ and ${\cal H}^{0} $ are constant matrices,
the first two being symmetric.    Then
\be {\tt H}=-\frac12\int dx ({\cal F}^0_{ij}v^i v^j+
{\cal G}^{0ij}v_i v_j +2{\cal H}^{0\;\;j}_{\;\;i} v^iv_j ) \;,
 \label{quadH} \ee  and using (\ref{voG}) or (\ref{voH}),
 we can express the Hamiltonian  on
either $LT^*G$ or $LT^*H$ as follows:
 \beqa {\tt H}(g,J^i)&=&
 -\frac12\int dx \biggl({\cal F}(g)_{ij}(\partial_x gg^{-1})^i
 (\partial_x gg^{-1})^j+
{\cal G}(g)^{ij}J_i J_j +2{\cal H}(g)^{\;\;j}_{i}
(\partial_x gg^{-1})^iJ_j \biggr) \;, \cr & &\label{quadHoG} \\
 {\tt H}(h,\tilde J_i)&=&
-\frac12\int dx \biggr(\tilde {\cal F}(h)_{ij}
\tilde J^i       \tilde J^j+\tilde
{\cal G}(h)^{ij} (\partial_x h h^{-1})_i (\partial_x h h^{-1})_j +  2
\tilde{\cal H}(h)^{\;\;j}_{i}
\tilde J^i (\partial_x h h^{-1})_j \biggl) \;,\cr & &
 \label{quadHoH} \eeqa
 respectively.  To evaluate  the functions
${\cal F}(g)$, ${\cal G}(g)$ and ${\cal H}(g)$
 we can first use (\ref{defab}) to write
 \beqa
  v^i& =& a(g^{-1})_j^{\;\;i}(\partial_x gg^{-1})^j  +
  b(g)^{ij} J_j\;,\cr
  v_i &=&a(g)_i ^{\;\;j} J_j  \label{vitoA}      \;.\eeqa
   Then we can substitute
into (\ref{quadH}) to find
\beqa      {\cal F}(g) &=& a(g)^{-1} {\cal F}^0\; { a(g)^{-1} }^T \cr
     {\cal H}(g) &=& a(g)^{-1} {\cal F}^0\;b(g) +  a(g)^{-1} {\cal H}^0
    \; a(g) \cr
  {\cal G}(g) &=& b(g)^T {\cal F}^0\;b(g) +  a(g)^T {\cal G}^0\; a(g)
+ b(g)^T  {\cal H}^0\; a(g)  +a(g)^T {  {\cal H}^0}^T \;
   b(g)       \;. \label{fgh}    \eeqa

Similarly, by writing
 \beqa   v^i &=&\tilde a(h)^i _{\;\;j}\tilde  J^j\;,\cr
  v_i& =& \tilde a(h ^{-1})^j_{\;\;i}(\partial_x hh^{-1})_j  +   \tilde
  b(h)_{ij} \tilde J^j   \label{vitoB}      \;,\eeqa
which follows from (\ref{deftab}),
$\tilde {\cal F}(h)$, $\tilde {\cal G}(h)$ and $\tilde {\cal H}(h)$
are given by:
\beqa
 \tilde {\cal F}(h) &=&\tilde  b(h)^T {\cal G}^0\;\tilde b(h) +
  \tilde  a(h)^T {\cal F}^0\;\tilde   a(h)
   +  \tilde b(h)^T { {\cal H}^0}^T\;\tilde  a(h)
      +\tilde a(h)^T {  {\cal H}^0} \; \tilde b(h)      \cr
 \tilde    {\cal H}(h) &=&\tilde b(h)^T {\cal G}^0\;{\tilde a(h)^{-1}}^T
  +\tilde  a(h)^{T} {\cal H}^0   \;  {\tilde a(h)^{-1}}^T  \cr
 \tilde {\cal G}(h)  &=& \tilde a(h)^{-1} {\cal G}^0\;
  {\tilde a(h)^{-1} }^T     \;. \label{tfgh}    \eeqa

In general, for an arbitrary choice
 of $R$, and hence the constant matrices
${\cal F}^0$, ${\cal G}^{0}$ and ${\cal H}^{0} $, the equations of motion
(\ref{defpld})  may not be
 Lorentz invariant.  On the other hand,  we can require that the system
 leads to familiar dynamics, thereby specifying $R$.
 Below we shall restrict to  the system of nonchiral massless
scalar fields, and thereby recover     known dynamical systems
possessing  T-duality.  We begin with the Abelian case.

\section{Abelian case}
\setcounter{equation}{0}

If both $G$ and $H$ are Abelian, the backgrounds
[${\cal F}(g)$, ${\cal G}(g)$, ${\cal H}(g)$]  and
[$\tilde {\cal F}(h)$, $\tilde {\cal G}(h)$, $\tilde {\cal H}(h)$]
reduce to the constant matrices
[${\cal F}^0$, ${\cal G}^{0}$, ${\cal H}^{0} $].  Furthermore, the
equations of motion (\ref{defpld}) reduce to
  \be   \partial_t \psi + R \partial_x \psi =0 \;,
       \label{abpld}  \ee
where $\psi=\phi^i e_i + \chi_i e^i$ is an element of the Lie algebra
of $D$.  Then
\beqa   \partial_t\chi_i &=&  {\cal H}^{0\;\;j}_{\;\;i} \partial_x\chi_j
+{\cal F}^0_{ij} \partial_x\phi^j   \cr
 \partial_t\phi^i&=&{\cal G}^{0ij} \partial_x\chi_j +
 {\cal H}^{0\;\;i}_{\;\;j}
 \partial_x\phi^j \;.\label{uafov}
 \eeqa
 For the simplest case of $n=1$,
these equations imply the existence of a pair of plane wave solutions
 with    velocities  determined by the three constants
  ${\cal H}^{0}$, ${\cal G}^{0}$ and ${\cal F}^{0}$
        \be
 [ \partial_t - ({\cal H}^{0} \pm \sqrt{{\cal G}^{0}{\cal F}^{0}})
   \partial_x] \;\psi^\pm =0\;,\qquad  \psi^\pm =
   \chi \pm \sqrt{{\cal F}^{0}/{\cal G}^{0}}\;\phi\;,\ee provided
${\cal G}^{0}\ne0$.   For arbitrary values of the constants
we will not get Lorentz invariant dynamics.
 To recover left and right moving  light-like
 waves  we  need ${\cal H}^{0}=0$ and
 ${\cal G}^{0}=1/{\cal F}^{0}$.  Eqs.
  (\ref{quadHoG}) and (\ref{quadHoH})
then reduce to the standard dual Hamiltonians for a massless scalar
field, with the duality transformation corresponding to the familiar
${\cal G}^{0}\rightarrow 1/{\cal G}^{0}$.\cite{kysy}
   [Alternatively, a chiral system
 with only right moving light-like waves
 results from the choice  ${\cal H}^{0}=1$
 and ${\cal G}^{0}={\cal F}^{0}=0$.  In this case, however, the system
is self-dual, as then the expressions for the Hamiltonians
(\ref{quadHoG})  and (\ref{quadHoH})    are identical.]

  For the  case where $n$ is arbitrary,
let us  demand    that (\ref{uafov})
  describes   $n$ nonchiral massless  fields.
Then $R$ should have eigenvalues $\pm 1$ with
eigenvectors of the form   $\psi=\phi^i e_i + \chi_i[\phi] e^i$,
 where $\chi_i$
 are determined from $\phi^i$, and $\phi^i$ are arbitrary (or vice versa).
  This is possible   provided that ${\cal G}^{0}$
is nonsingular and
 \be
   {\cal F}^{0} = ( {\cal H}^{0} + \BI) {{\cal G}^{0}}^{-1}
( {\cal H}^{0} + \BI)^T
    = ( {\cal H}^{0} - \BI) {{\cal G}^{0}}^{-1}
( {\cal H}^{0}- \BI)^T
\;.\ee       This then implies that the matrix  ${\cal B}^0=
{\cal H}^{0}{{\cal G}^{0}}^{-1} $ is antisymmetric.
We denote by   $\{T_\alpha^\pm$,  $i=1,2,...n$,   $\alpha=\pm\}$
the eigenvectors of $R$
  \be  R T^\alpha_i   =\alpha T^\alpha_i \;.\label{defR}\ee
They are given by
\be T^\pm_i = e_i -{\cal E}_{ij}^\pm  e^j   \;,\qquad
{\cal E}^{\pm}_{ij}=\pm{\cal G}^0_{ij}+{\cal B}^0_{ij} \;,    \label{Tinoe} \ee
where ${\cal G}^0_{ij}$ are the matrix elements of $ {{\cal G}^{0}}^{-1}$,
and   have inner product
$ <T^\alpha_i|T^\beta_j>=-2\alpha\delta_{\alpha,\beta}
 {\cal G}^0_{ij}\;.$
 By substituting the above result for the constant matrices, i.e.
\beqa   {\cal F}^0_{ij} &=&  {\cal G}^0_{ij} - {\cal B}^0_{ik}
{\cal G}^{0k\ell}  {\cal B}^0_{\ell j}  \cr
{\cal H}^{0\;\;j}_{\;\;i} &=&  {\cal B}^0_{ik} {\cal G}^{0kj}  \;,
\label{sofHF}    \eeqa     into   (\ref{uafov}),
 we get the standard equations of motion for  massless modes  familiar
in string theory\cite{narain}\cite{gsw} and also in the quantum Hall
effect\cite{bal}.

  In addition to the two dual descriptions, described in the previous
section, it is well known (in the Abelian case) that
 there is an entire family of dual descriptions related by
$O(n,n)$ canonical transformations\cite{shwi}\cite{bal}:
\be \pmatrix{v^i \cr v_i } \rightarrow {\cal O}  \pmatrix{v^i \cr v_i }
\;,\qquad  {\cal O}  \pmatrix{& \BI\cr\BI &} {\cal O}^T=
\pmatrix{& \BI\cr\BI &}  \;.\ee
  Topological and quantum considerations further restrict
the transformations to a discrete subgroup corresponding to
$O(n,n,Z)$.       Under such transformations, the constant
matrices  ${\cal H}^{0}$, ${\cal G}^{0}$ and ${\cal F}^{0}$
in the Hamiltonian are transformed by  \be
\pmatrix{{\cal F}^{0}&{\cal H}^{0} \cr {{\cal H}^{0} }^T &{\cal G}^{0}}
 \rightarrow  {\cal O}^T
\pmatrix{{\cal F}^{0}&{\cal H}^{0} \cr {{\cal H}^{0} }^T &{\cal G}^{0}}
 {\cal O}   \;.\ee

 \section{Nonabelian case}
\setcounter{equation}{0}

 The nonabelian generalization of Abelian T-duality
 considered by Klimcik and Severa\cite{ks95} utilizes
  the same definition of $R$ as       in the Abelian case,  i.e.
(\ref{defR}) and (\ref{Tinoe}).
At the level of the equations of motion, duality was achieved
  after projecting  (\ref{defpld}) onto
  $G$ and $H$.     We now review the procedure.

  By parametrizing the Drinfeld double variables according to
$\ell=\tilde h g$, $\tilde h\in H$ and $g\in G$, one gets
\be \ell^{-1} d\ell=g^{-1} \tilde h^{-1}d \tilde hg +g^{-1} d g\;.\ee
Comparing with (\ref{voG}) we see that $J_ie^i$ is identified with
$\tilde h^{-1}\partial_x \tilde h$.  Eq.
(\ref{defpld}) can be expressed as
\be
 <T^\pm_i|\ell^{-1}\partial_\pm\ell > =0\;,\qquad \partial_\pm=
\partial_t\pm \partial_x             \ee and hence \be
<gT^\pm_ig^{-1}|A_{\pm i}e^i+A_\pm^ie_i > =0 \;,\label{chir2}\ee
               where $A_\pm^ie_i=\partial_\pm gg^{-1}$,
 $ A_{\pm i}e^i=\tilde h^{-1}\partial_\pm\tilde h$,
  and we  used the invariance property of the inner product.
To evaluate      (\ref{chir2}), we can use
(\ref{defab}) and express  $gT^\pm_ig^{-1}$    according to \beqa
  gT^+_ig^{-1}&=& M^{+}(g)_i^{\;\;j}[  e_j - E(g)_{jk} e^k ]  \cr
 gT^-_ig^{-1}&=& M^{-}(g)_i^{\;\;j}[  e_j + E(g)_{kj} e^k ]\;,\eeqa
    where
 \be  M^\pm(g)= a(g) -{\cal E}^\pm b(g)   \;, \ee  and
 \be E(g)= { M^+(g)}^{-1} {\cal E}^+ {a(g)^T}^{-1}\;.\label{Eafog}\ee
Equations (\ref{chir2}) become
        \beqa   A_{+ i} - E(g)_{ij}A^j_+&=&0 \cr
         A_{- i} + E(g)_{ji}A^j_-&=&0 \;.\label{Apm}\eeqa
These equations define the  connection components
$ A_{\pm i} $  as functions on  $TG$.  They satisfy
the Maurer-Cartan equations associated with $H$,
\be
\partial_- A_{+i} -  \partial_+ A_{-i} +c_i^{jk} A_{j-} A_{k+} =0 \;.
\label{mce1}  \ee
                  These equations were obtained
 from an action principle written on $TG$, the Lagrangian density being
\be L  = E(g)_{ij}  A^i_- A^j_+   \;.\label{loG}\ee    Eq.
 (\ref{mce1}) results from extremizing $\int d^2x L$ in $g$,
  using the definition
(\ref{Eafog}) of the background $E(g)$.  In addition, we have the
 Maurer-Cartan equations associated with $G$,
 \be \partial_- A_+^i -  \partial_+ A_-^i -c^i_{jk} A^j_- A^k_+ =0 \;,\ee
 which here play the role of identities.
The  Hamiltonian associated with the  Lagrangian
(\ref{loG}) is written on $LT^*G$ and was derived in \cite{Sfet}.  It is
just   (\ref{quadHoG})   with constant
matrices given by (\ref{sofHF}).

The dual description is obtained by   instead
  factorizing the Drinfeld double variables according to
 $\ell=\tilde g    h$,   $\; h\in H$ and $\tilde g\in G$, and
repeating the above procedure.
Now from (\ref{voH}), $\tilde J^ie_i$ is identified with
$\tilde g^{-1}\partial_x \tilde g$.
 Using (\ref{deftab}), the analogue
of the matrix $E(g)$ is
\be\tilde E ( h) =[\tilde a ( h) -{{\cal E}^+}^{-1}\;\tilde b
( h) ]^{-1} {{\cal E}^+}^{-1} \;{\tilde a ( h)^T}^{-1}
  \;.\label{deftE}\ee
  Defining
$\tilde  A_{\pm i}e^i=-\partial_\pm h h^{-1}$ ,
 $\tilde  A_\pm^ie_i=-\tilde g^{-1}\partial_\pm \tilde g$,
the  equations (\ref{chir2}) can now be written as
  \beqa \tilde A^i_+ -\tilde  E( h)^{ij}\tilde A_{+ j}& =&0\cr
\tilde A^i_- +\tilde  E( h)^{ji}\tilde A_{- j}& =&0 \label{tApm}
  \eeqa
stating that the  connection components    $ \tilde A^i_\pm   $  are
 functions on  $TH$.  They satisfy
the Maurer-Cartan equations associated with $G$,
\be \partial_-\tilde A_+^i -
 \partial_+ \tilde A_-^i -c^i_{jk}\tilde A^j_
- \tilde A^k_+ =0 \;.\label{mce2}\ee
  This is obtainable from
an action principle written on $TH$, the Lagrangian density being
\be\tilde L  =\tilde E( h)^{ij} \tilde A_{i-} \tilde A_{j+}
\label{loH}  \;.\ee  Eq. (\ref{mce2}) results from extremizing
 $\int d^2x \tilde L$   in $ h$, using the definition (\ref{deftE}) of
 the background  $\tilde E ( h)$.
      In addition, we    have the Maurer-Cartan equations associated
with $H$
\be\partial_- \tilde A_{+i} -  \partial_+ \tilde A_{-i}
+c_i^{jk} \tilde A_{j-} \tilde A_{k+} =0 \;, \ee  which
now are identities.
Thus the roles of identities and equations of motion are reversed in
the two descriptions.
The  Hamiltonian associated with the  Lagrangian
(\ref{loH}) is written on $LT^*H$ and  is
just   (\ref{quadHoG})   with constant
matrices again given by (\ref{sofHF}).

 \section{$O(3)$ nonlinear $\sigma$-model}
\setcounter{equation}{0}

We have argued that Poisson Lie T-duality need not be tied to any
particular dynamics on $G$ or $H$.  By making relatively simple
assumptions on the Hamiltonian, we have shown how to
 recover the known dynamical  systems having Poisson Lie T-duality.
  But alternative
dynamics  should also be of interest.   As an example, here
we consider  modifying
  the dynamics by introducing gauge symmetries.  This will
 allow us to describe coset models.  Here we will just examine
the simple case of the $O(3)$ nonlinear $\sigma$-model.

In the Hamiltonian formulation we can  introduce gauge symmetries
 by imposing
first class constraints.  The latter should generate a
subgroup of $\widehat{LD}$.  For example, one can set one component of
$v$ (weakly) equal to zero, say $v_1$.  The resulting gauge invariant
physical subspace will be $2n-2$ dimensional.  Upon projecting the
 constraint    onto $LT^*G$, we get
 \be \Phi(g,J_i)=a(g)^{\;\;i}_1 J_i\approx 0 \;,\label{coTG}\ee
  generating right transformations
on $G$:    $  \delta g =   \epsilon g e_1$ ,
  $\epsilon$ being infinitesimal.
On $LT^*H$ the constraint will look like
 \be \tilde\Phi(h,\tilde J^i)= \tilde b(h)_{1i} \tilde J^i +
[\tilde a(h^{-1})]^i_{\;\;1} (\partial_x h h^{-1})_i \approx 0 \;,
\label{coTH}\ee
generating a more complicated set of  orbits on $H$:
$  \delta h =  \epsilon   \tilde b(h)_{1i}  e^i h$ .

Let us  now specialize to the case where $D=T^*SU(2)$, with $G=SU(2)$ and
$H$ is the three dimensional  Abelian group.
  This has been referred to as the semiabelian case.  Then
$c^k_{ij}=\epsilon_{ijk}$ and $c_k^{ij}=0$, $i,j,k,..=1,2,3$.
   From (\ref{defab}) and
(\ref{deftab}) it
follows that \be b(g)^{ij}=0\;,\qquad \tilde a(h)^i_{\;\; j}=\delta^i_j
\;, \qquad \tilde b(h)_{ij}  =\epsilon_{ijk} \chi_k \;,\ee
where we write $h=\exp(\chi_i e^i)$.        Once
we impose the constraint (\ref{coTG}) on $LT^*G$ we are left with
four gauge invariant field degrees of freedom.
 The corresponding  configuration
space is spanned by fields $\psi^i(x)$ having values in $SU(2)/U(1)$,
  the target  space for the
$O(3)$ nonlinear $\sigma$-model.
$\psi^i(x)$ satisfy the constraint $\psi^i(x)\psi^i(x) =1$, and can be
  defined by
\be      \psi^i(x)e_i= g(x)e_1 g(x)^{-1} \;,\ee  $g(x)$  having values
in $SU(2)$.  The standard Lagrangian density is\cite{book}
\be L= \frac12 \partial_\mu \psi^i   \partial^\mu\psi^i =
\frac12 ( g^{-1}\partial_\mu g)_a   (g^{-1}\partial^\mu g)_a \;,
\qquad      a=2,3    \;,
 \ee  where $\mu$ is the space time index.
    It is straightforward to derive
the corresponding canonical Hamiltonian.  Up to a Lagrange multiplier
term  involving the constraint (\ref{coTG}), it can be written
\be {\tt H}(g,J^i)=
  -\frac12\int dx \biggl([a(g)^{-1}]_i^{\;\;a} [a(g)^{-1}]_j^{\;\;a}
  (\partial_x gg^{-1})^i (\partial_x gg^{-1})^j+
 [a(g)]_a^{\;\;i} [a(g)]_a^{\;\;j}
  J_i J_j    \biggr) \;. \label{sigH} \ee $J_i$ generate left
   transformations on $G$ as in (\ref{ltoG}), and there are no
secondary constraints.
Eq. (\ref{sigH}) is consistent with having a quadratic Hamiltonian on
$\widehat {LD}$ and projecting onto $LT^*G$.       Comparing with
  (\ref{quadHoG})  and (\ref{fgh}),  we can identify the constant
matrices [${\cal F}^0$, ${\cal G}^{0}$, ${\cal H}^{0} $] for this case.
  The  only nonzero components are
 \be
 {\cal F}^0_{22}={\cal F}^0_{33}=1 \;,\qquad
 {\cal G}^{022}  = {\cal G}^{033}=1\;. \ee
From (\ref{deftab}) and  (\ref{tfgh}) we can then read off the dual
Hamiltonian.  In terms of the canonically conjugate variables
$\chi_i$ and $\tilde J^i$ it is given by: \be
 {\tt H}(h,\tilde J^i) =
-\frac12\int dx \biggr(\tilde J^a   \tilde J^a +
(\epsilon_{aij} \tilde J^i\chi_j  +  \partial_x \chi_a  )^2  \biggl) \;,
\label{dho3}  \ee
up to the  constraint  (\ref{coTH}), which  here  reduces to
 \be \tilde\Phi(h,\tilde J^i)= \epsilon_{1ab} \tilde J^a\chi_b +
\partial_x \chi_1 \approx 0 \;.\label{dcon}   \ee
Eq. (\ref{dho3}) differs from the dual Hamiltonian of the principal
chiral model\cite{loz} since the index $a$ is restricted to values
$2,3$.   From (\ref{dcon}) gauge transformations are associated with
rotations about the $1-$axis in the target space spanned by $\chi_i$.

Lastly, we note that the above dual formulation of the
$O(3)$ nonlinear $\sigma$-model is not unique.  This is because
there exists at least one other Drinfeld double group $D$ containing
containing $SU(2)$ as a maximally isotropic subgroup.  It is $D=SL(2,C)$.
Here one can take $H=SB(2,C)$.\cite{bss}
The above procedure can be repeated in this case to find a dual
Hamiltonian formulation
of the $O(3)$ nonlinear $\sigma$-model written on $LT^*SB(2,C)$.

   \section*{Acknowledgments}

I am grateful to S. Frolov, G. Marmo, F. Lizzi, A. Simoni and P. Vitale
 for the stimulating discussions.  I thank G. Marmo
and members of Dipartimento di Scienze Fisiche, Universit\`a di
Napoli, for their  hospitality and support where this work was
initiated.  A.S. was  supported in part by the U.S. Department of
Energy under  contract number DE-FG05-84ER40141.

    \end{document}